\documentclass[twocolumn,pra]{revtex4}
\usepackage[utf8]{inputenc}
\usepackage[english]{babel}
\usepackage{amsmath}
\usepackage{amsfonts}
\usepackage{amssymb}
\usepackage{lmodern}
\usepackage{esint}
\usepackage{braket}
\usepackage{color}
\usepackage{array}

\begin{document}
\preprint{NT@UW-15-11}
\title{On the proton charge extensions}

\author{Jesse R. Stryker}
\email{stryker@uw.edu}

\author{Gerald A. Miller}
\email{miller@uw.edu}

\affiliation{Department of Physics, University of Washington, Seattle, Washington}

\date{January 14, 2016}

\begin{abstract}
We examine how corrections to $S$-state energy levels, $ E_{nS}$, in hydrogenic atoms due to the finite proton size are affected by moments of  the proton charge distribution.
The corrections to $E_{nS}$ are computed moment by moment. The results demonstrate that the next-to-leading order term in the expansion is of order $r_p / a_B $ times the size of the leading order $ \langle r_p^2 \rangle $ term.
Our analysis thus dispels any concern that the larger relative size of this term for muonic hydrogen versus electronic hydrogen might account for the current discrepancy of proton radius measurements extracted from the two systems.
Furthermore, the next-to-leading order term in powers of $r_p / a_B $ that we derive from a dipole proton form factor is proportional to $\langle r_p^3 \rangle $, rather than $\langle r_p^4 \rangle$ as would be expected from the scalar nature of the form factor.
The dependence of the finite-size correction on $\langle r_p^3 \rangle $ and higher odd-power moments is shown to be a general result for any spherically symmetric proton charge distribution.
A method for computing the moment expansion of the finite-size correction to arbitrary order is introduced and the results are tabulated for principal quantum numbers up to $n=7$.
\end{abstract}


\maketitle

\renewcommand{\vec}[1]{\textbf{#1}} 

\section{Introduction}
The proton radius puzzle concerns the incompatibly different measurements of the proton radius $r_p\equiv \langle r_p^2 \rangle^{1/2} $ resulting from normal hydrogen ($ep$) spectroscopy and spectroscopy of muonic hydrogen ($\mu p$).
The 2010 recommended CODATA value based on hydrogen spectroscopy and electron-proton scattering is $r_p = 0.8775(51)$ fm, \cite{RevModPhys.84.1527} whereas the CREMA collaboration found $r_p = 0.84087(39)$ fm from precision measurements of the Lamb shift in $\mu p$.  \cite{Pohl:2010,Antognini25012013}
In atomic spectroscopy, the size of $ \langle r_p^2 \rangle $ is calculated from its shift of the $S$-state energy levels, which is given to leading order in powers of the proton radius by
\begin{equation}
  \Delta E_{nS} = \frac{2\pi \alpha}{3} \left| \psi_{nS} (0) \right|^2 \langle r_p^2 \rangle = \frac{2 \alpha }{3 n^3 a_B^3 } \langle r_p^2 \rangle .
  \label{eq:LOcorrection}
\end{equation}
In scattering experiments, $r_p$ is obtained from the Sachs  electric form factor $G_E(Q^2)$ according to  
\begin{equation}
  \langle r_p^2 \rangle = -6 \left. \frac{d G_{E}(Q^2 )}{d (Q^2 )} \right|_{Q^2 = 0} .
  \label{eq:rpdefinition}
\end{equation}
Reviews of the proton radius puzzle by R. Pohl \emph{et al.} \cite{Pohl:2013yb} and C. Carlson \cite{Carlson:2015jba} discuss in detail the importance of this problem and include various theoretical explanations that have yet to gain acceptance.

Among the explanations of any discrepancy between theory and experiment, there must always be included the possibility that the theory was not correctly applied, and in this article we investigate one particular possibility.
% the question of whether or not higher-order contributions to Eq. (\ref{eq:LOcorrection}) are, in fact, negligible, as they have been regarded thus far.
We note that 
Eq. (\ref{eq:rpdefinition}) arises from using  the Sachs electric form factor $G_E(Q^2)$ in a first-order expansion in terms of  $Q^2=\vec{q}^2$.
As pointed out in \cite{Gluck:2014qxa}, the proton radius puzzle makes it natural to consider the effects of the $\vec{q}^4$ and higher-order  terms in this expansion.
Retaining such terms in the   form-factor and proceeding as with the first-order calculation leads to the  proposal of a lepton-dependent parametrization of the leading order finite-size energy correction to the $nS$ levels of hydrogenic atoms,
\begin{equation}
  \Delta E_{nS,\ell}=\frac{2\pi \alpha}{3} | \psi_{nS,\ell}(0) |^2 r_{p,\ell}^2 ,
  \label{eq:gluckdeltaE}
\end{equation}
where $\ell \in \{ e, \mu \} $.
Since the muonic Lamb shift measured by the CREMA collaboration concerns a $2S$-$2P$ transition, we demonstrate that the proposed parametrization~\cite{Gluck:2014qxa} of $r_{p,\ell}^2$ leads to a divergence in the energy correction of Eq.\ (\ref{eq:gluckdeltaE}). 
We go on to examine the argument of \cite{Gluck:2014qxa}    based on continuing the expansion of Eq.\ (\ref{eq:LOcorrection}) in powers of $\vec{q}^2$  and confirm that the next term in the series is indeed more significant for $\mu p$ than for $ep$.
However, as will be shown, the effect remains far too small to account for a 4\% discrepancy in $r_p$.
Our analysis parallels the work of \cite{Carroll:2011de,Carroll:2011rv},  which  reached the same conclusions using numerical methods, and we generalize those results.
\section{Divergence of the proposed corrections}

The proposed lepton-dependent radius extracted from $nS$ energy levels which appears in Eq. (\ref{eq:gluckdeltaE}) arises from expanding $G_E$ in powers of $\vec{q}^2$  \cite{Gluck:2014qxa}, an idea that is well-motivated.
Keeping terms up to $\vec{q}^4$ and rearranging the result so as to match the factorization in Eq. (\ref{eq:gluckdeltaE}) would seem to suggest the reparametrization,
\begin{align}
  r_{p,\ell}^2 & = r_p^2 - \langle \vec{q}^2 \rangle_{n,\ell} \tilde{r}_{p}^4 , \label{} \label{eq:gluckrp2} \\
  \langle \vec{q}^2 \rangle_{n,\ell} & = |\psi_{nS,\ell}(0)|^{-2} \int \frac{d^3 q}{(2\pi )^3 } \vec{q}^2 \int d^3 r e^{i \vec{q}\cdot \vec{r}} |\psi_{nS,\ell}(\vec{r})|^2 , \label{eq:gluckq2}
\end{align}
with $\tilde{r}_p^4$ a parameter determined by the second derivative of $G_E$.
An explicit expression for $\tilde{r}_{p}^4$ will not be needed.
Let us now consider as an example the $2S$ wavefunction, which  is well-known to be
\begin{equation}
  \psi_{2S,\ell }(\vec{r}) = \frac{1}{\sqrt{4\pi }} \frac{1}{\sqrt{2 a_\ell^3 }} \left( 1-\frac{r}{2 a_\ell } \right) e^{-\tfrac{r}{2 a_\ell  }}
  \label{eq:wavefn}
\end{equation}
where
\begin{equation}
  a_{\ell} \equiv\frac{1}{\alpha m_{\text{red},\ell} }
  \label{}
\end{equation}
denotes the Bohr radius of either $e p$ or $\mu p$, and we work in center-of-mass coordinates.
Inserting this wavefunction into Eq. (\ref{eq:gluckq2}) yields
\begin{equation}
  \begin{split}
    \langle \vec{q}^2 \rangle_{2,\ell} =&  |\psi_{2S,\ell}(0)|^{-2} \int \frac{d^3 q}{(2\pi )^3 } |\vec{q}|^2 \\
    & \times \int d^3 r\, e^{i \vec{q}\cdot \vec{r}} \frac{1}{8\pi a_\ell^3 } \left( 1-\frac{r}{2 a_\ell } \right)^2 e^{-r/a_\ell } .
  \end{split}
\end{equation}
Evaluating the q-space integral first leads to an immediate divergence.
Equivalently, using the fact that $ |\vec{q}|^2 e^{ i \vec{q} \cdot \vec{r}} = - \nabla^2 e^{i \vec{q} \cdot \vec{r} }$ and integrating by parts, one finds that this divergence corresponds to trying to evaluate $\nabla^2 | \psi_{2S,\ell } (\vec{r} ) |^2 $ at the origin, which can be seen to diverge using the Schr\"{o}dinger equation.
These considerations, which hold for any $S$-state, show the proposed finite-size correction is divergent, so the idea given by \cite{Gluck:2014qxa} cannot bring the different measurements of $r_p$ into agreement.
Nevertheless, it is worthwhile to pursue the idea that higher $\vec{q}^{2n}$ terms might be important for the finite-size correction in muonic hydrogen.
This is done in the following by first evaluating the relevant matrix element as a function of $G_E$ and then expanding the result in terms of the ratio of the size of the proton to the Bohr radius.
This calculation will therefore show that using the proper order of integration and series expansion is essential to getting the correct result.

\section{Expansion of the $S$-state finite-size correction}
We now examine what the true effect of higher-order moments is on the size of the $nS$-level corrections derived from nonrelativistic (NR) perturbation theory.
\subsection{Calculation from an approximate NR proton form factor}
The perturbation to the Hamiltonian is given in coordinate space by
\begin{equation}
  \delta V (\vec{r} ) = -\alpha \int d^3 r^{\prime } \left[ \frac{\rho (\vec{r}^\prime )-\delta (\vec{r}^\prime ) }{| \vec{r}-\vec{r}^\prime |} \right] ,
  \label{eq:positionperturbation}
\end{equation}
where $\rho $ describes the electric charge distribution of the proton (normalized to unity).
The corresponding energy correction to the $nS$ level can be written as
\begin{equation}
  \delta E_{nS}= \int d^3 r | \psi_{nS} (\vec{r}) |^2 \int \frac{d^3 q}{(2\pi )^3 } e^{i \vec{q}\cdot \vec{r} } \delta V (\vec{q})
  \label{eq:correctionFT}
\end{equation}
where $\delta V (\vec{q})$ is the Fourier transform of Eq.\ (\ref{eq:positionperturbation}), 
\begin{equation}
  \delta V (\vec{q}) = \frac{4\pi\alpha}{q^2} \left[ 1- F (\vec{q}) \right] ,
  \label{eq:qspaceperturbation}
\end{equation}
with $F(\vec{q})$ being the NR form factor.
To proceed with analyzing the higher-order terms in $\delta E_{nS}$ we first approximate the form factor with what is traditionally known as the dipole form factor:
\begin{equation}
  G_E(\vec{q})=  F_{\text{dipole}} (\vec{q}) = \left( 1+\frac{|\vec{q}|^2}{ \Lambda^2} \right)^{-2},
  \label{eq:dipoleform}
\end{equation}
where $\Lambda^2 \simeq 0.71\,\text{GeV}^2$. The name dipole arises from the fact that $G_E$ has a second-order pole at $|\vec{q}|^2=-\Lambda^2$. 
Substituting the dipole form factor into Eq.\ (\ref{eq:qspaceperturbation}) and evaluating the q-space integral of Eq.\ (\ref{eq:correctionFT}) leads to
\begin{equation}
  \delta E_{nS} = \alpha \int_0^\infty dr\, r^2 \left| R_{n0} (r) \right|^2 \left( \frac{\Lambda r +2}{2} \right) \frac{e^{-\Lambda r}}{r}  .
  \label{}
\end{equation}
In particular, one finds that
\begin{equation}
  \delta E_{2S} = \alpha\Lambda \left[ \frac{4 (a_\ell\Lambda )^{-3}+3(a_\ell\Lambda )^{-5} + (a_\ell\Lambda )^{-6} }{4 (1+(a_\ell\Lambda )^{-1} )^5} \right].
  \label{}
\end{equation}

Since $ (a_\ell \Lambda )^{-1} \ll 1$ for both $e p$ and $\mu p$, this quantity can be used as an expansion parameter:
\begin{equation}
  \begin{split}
    \delta E_{2S}  = \alpha\Lambda & \left[ (a_\ell \Lambda )^{-3} -5 (a_\ell \Lambda )^{-4} +\frac{63}{4}(a_\ell \Lambda )^{-5} \right.\\
    & \left. -\frac{77}{2} ( a_\ell \Lambda )^{-6}+ 80 (a_\ell \Lambda )^{-7} - \cdots \right] .
  \end{split}
  \label{eq:Lambdaseries}
\end{equation}
It will be economic to define $\xi_{nS}({\rho ,\ell})$ as the ratio of the second term to the first term in an expansion of the form in Eq. (\ref{eq:Lambdaseries}) for $\delta E_{nS}$, resulting from the distribution $\rho$ with satellite $\ell$.
In order to account for a 4\% reduction in the apparent $r_p$ (an 8\% reduction in $r_p^2$), we require $\xi_{2S}({\rho ,\mu})\approx 8\%$.
Using the known values of $a_\ell $ and $\Lambda$, we calculate
\begin{equation}
  \xi_{2S}({\rho ,\ell}) \approx
  \begin{cases}
    -4.1\times 10^{-3}, & \ell = \mu, \\
    -2.2\times 10^{-5}, & \ell = e,
  \end{cases}
  \label{eq:sizeratio}
\end{equation}
for a dipole form factor.
The relative size $\xi_{2S}(\rho ,\mu )=-4.1\times 10^{-3}$ therefore shows that neglecting higher-order moments in Eq.\ (\ref{eq:LOcorrection}) is not to blame for the proton radius discrepancy [despite the fact that $\xi_{nS}(\rho , \ell)$ is about 200 times greater for $\mu p$ versus $e p$].
More concretely, as discussed in \cite{Pohl:2010}, any missing terms from the energy difference associated with the muonic Lamb shift would need to amount to $0.31$ meV in order to bring their value of $r_p$ into agreement with the previously accepted CODATA value, and the second term of Eq.\ (\ref{eq:Lambdaseries}) amounts to only a $0.02$ meV difference.
The conclusions we have reached so far, namely the inadequacy of higher-order terms in the finite-size correction as a solution to the proton radius puzzle, will next be substantiated without assuming a specific shape of the proton form factor.
\subsection{General dependence of $E_{2S}$ on moments of the proton charge distribution}
In the preceding section, the assumption of a dipole form factor lead to the expansion in Eq.\ (\ref{eq:Lambdaseries}) for $\delta E_{2S}$.
However, Eq.\ (\ref{eq:Lambdaseries}) can be cast in a more revealing form by trading the powers of $\Lambda $ for moments of the proton charge distribution,
\begin{equation}
  \langle r_p^n \rangle = \int \rho(\vec{r}) r^n\, d^3r.
  \label{eq:momentdef}
\end{equation}
The dipole form factor corresponds to a proton distribution function,
\begin{equation}
  \rho (\vec{r}) = \Lambda^3 e^{-\Lambda r} / 8\pi ,
  \label{eq:protondensity}
\end{equation}
from which one finds that
\begin{equation}
  \Lambda^{-n} = 2 \langle r_p^{n} \rangle / (n+2)! .
  \label{eq:lambdatomoments}
\end{equation}
Thus, the expansion in Eq.\ (\ref{eq:Lambdaseries}) reads
\begin{equation}
  \begin{split}
    \delta E_{2S}/\alpha = & \frac{ \langle r_p^2 \rangle }{12 a_{\ell}^3 } - \frac{ \langle r_p^3 \rangle }{12 a_{\ell}^4 } + \frac{ 7 \langle r_p^4 \rangle }{160 a_{\ell}^5 }\\
    & - \frac{11 \langle r_p^5 \rangle }{720 a_\ell^6 } + \frac{ \langle r_p^6 \rangle }{ 252 a_\ell^7 } - \cdots .
  \end{split}
  \label{eq:momentseries}
\end{equation}

At this point we call attention to the peculiar result that all moments of $r_p$ higher than the first appear in the finite-size correction Eq.\ (\ref{eq:momentseries}).
It is straightforward to show that the form factor of a spherically symmetric density is given by
\begin{equation}
  F(\vec{q})|_{\rho(\vec{r} ) = \rho (|\vec{r}| ) } = \sum_{m=0}^\infty \frac{ (-)^m \langle r^{2m} \rangle }{(2m+1)!} q^{2m},
  \label{eq:formfactorexpansion}
\end{equation}
i.e., the form factor can be written purely in terms of even moments of the coordinate-space density.
In light of Eq..\ (\ref{eq:correctionFT}) and (\ref{eq:qspaceperturbation}), the appearance of odd-power moments in Eq.\ (\ref{eq:momentseries}) is surprising.
We believe that the form of the series in Eq.\ (\ref{eq:formfactorexpansion}) may have been the motivation for the lepton-dependent parametrization of Eq.\ (\ref{eq:gluckrp2}).
We note that the term proportional to $ \langle r_p^3 \rangle $ in Eq.\ (\ref{eq:momentseries}) is not at all related to the
third Zemach moment~\cite{Friar:2005jz,Cloet:2010qa}, which produces a much larger change to the energy level.

Although we obtained the expansion of Eq.\ (\ref{eq:momentseries}) by assuming a dipole form factor, this is in fact the general result for any spherically symmetric proton charge density.
Other models for the density which can be verified to yield the same expansion as Eq.\ (\ref{eq:momentseries}) include
\begin{align}
  \rho_{\text{G}}(\vec{r}) & = \left[ \frac{3}{2\pi \langle r_p^2 \rangle } \right]^{3/2} \exp \left( - \frac{3 r^2 }{2 \langle r_p^2 \rangle } \right) , \\
  \rho_{\text{Y}}(\vec{r}) & = \frac{3}{2\pi \langle r_p^2 \rangle r} \exp \left(  \frac{-\sqrt{6} r }{ \langle r_p^2 \rangle^{1/2} } \right)  ,
  \label{}
\end{align}
although it should be noted that the values of the moments themselves depend, of course, on which density is used.
(Densities of the above forms were also used in the numerical work of \cite{Carroll:2011de}.)

To justify our result in general, we claim that the spherically symmetric density appearing in Eq.\ (\ref{eq:positionperturbation})---whatever its particular form may be---can be represented formally by the series
\begin{equation}
  \rho (\vec{r}) \, \dot{=} \, \frac{1}{4\pi r^2 }\sum_{j=0}^{\infty} \frac{ (-)^j \langle r_{p}^{j} \rangle }{j!} \delta^{(j)}(r), 
  \label{eq:densityseries}
\end{equation}
where $\int_0^\infty \delta (r) dr = 1$ and $\delta^{(j)}(r)\equiv(\frac{d}{dr})^j \delta(r) $.
This allows us to express the finite-size correction to the $nS$ level as
\begin{equation}
  \delta E_{nS} = -\alpha \sum_{j=1}^{\infty} \int dr^\prime \, {r^\prime}^2 | R_{n0} (r^\prime ) |^2 \int dr\, \frac{ (-)^j \langle r_{p}^{j} \rangle }{j!\, r_> } \delta^{(j)}(r)
  \label{eq:energyseries}
\end{equation}
where $r_> \equiv \max (r,r^\prime )$.

It is not immediately obvious how one should go about evaluating the terms in Eq.\ (\ref{eq:energyseries}) since $\delta (r)$ lives on the non-negative real axis rather than the usual entire real line;
typical methods such as integration by parts lead to ambiguous expressions involving, e.g., $\delta (0)$ or $\delta^\prime (0)$.
One solution in such circumstances is to replace the Dirac delta with a representative test distribution $f(r,\epsilon )$ which becomes $\delta (r)$ in the limit $\epsilon \rightarrow 0$.
However, many of the usual test functions diverge for terms $j>2$ when the $\epsilon \rightarrow 0$ limit is taken.

We find that the following sequence of test functions can reliably be used to evaluate any term of the series in Eq.\ (\ref{eq:energyseries}):
\begin{equation}
  f_j(r,\epsilon)=\frac{ e^{-r/\epsilon } }{(j+2)! \epsilon } \left( \frac{r}{ \epsilon } \right)^{j+2} .
  \label{eq:deltamodel2}
\end{equation}
These test distributions are unit-normalized on the positive real axis for any finite $\epsilon >0$ and the $j$th test function is used to evaluate the $j$th term of Eq.\ (\ref{eq:energyseries}).
In contrast to the usual prototypes used for the entire real line, our test functions peak on the positive real axis and only reach the origin as $\epsilon \rightarrow 0$.

  \begin{table*}[t]
    \caption{\label{tab:momentcoeffs} Numerical coefficients $c_n^{(j)}$ of the moment expansion for $\delta E_{nS} $ (cf. Eq.\ (\ref{eq:momentseries2ndderivation})). The coefficients are calculated according to the method described in section \ref{sec:momentdep}. The $j$=1 column of zeroes is just the restatement that the finite-size correction to $E_{nS}$ has no dependence on $\langle r_p \rangle$. The $j=2,3$ columns suggest that $c_n^{(3)}=-c_n^{(2)}$ for all $n$.}
      \def\arraystretch{1.4}
      \setlength\tabcolsep{2.5mm}
    \begin{tabular}{lcccccccccc}
      \hline
      \hline
  $c_n^{(j)}$  & $j$=1 & $j$=2 & $j$=3 & $j$=4 & $j$=5 & $j$=6 & $j$=7 & $j$=8 & $j$=9 & $j$=10\\ \hline
  $n$=1  & 0 & $\tfrac{2}{3}$ & $\tfrac{-2}{3}$ & $\tfrac{2}{5}$ & $\tfrac{-8}{45}$ & $\tfrac{ 4}{ 63}$ & $\tfrac{-2 }{105 }$ & $\tfrac{2 }{405 }$ & $\tfrac{-16 }{14175 }$ & $\tfrac{4}{17 325}$\\ 
  $n$=2  & 0 & $\tfrac{1}{12}$ & $\tfrac{-1}{12}$ & $\tfrac{7}{160}$ & $\tfrac{-11}{720}$ & $\tfrac{ 1}{252 }$ & $\tfrac{-11 }{13440 }$ & $\tfrac{ 29}{ 207360}$ & $\tfrac{ -37}{1814400 }$ & $\tfrac{23}{8 870 400}$\\
  $n$=3  & 0 & $\tfrac{2}{81}$ & $\tfrac{-2}{81}$ & $\tfrac{46}{3645}$ & $\tfrac{-136}{32805}$ & $\tfrac{404 }{413343 }$ & $\tfrac{-122 }{688905 }$ & $\tfrac{ 206}{ 7971615}$ & $\tfrac{ -7888}{ 2511058725}$ & $\tfrac{3004}{9 207 215 325}$\\
  $n$=4  & 0 & $\tfrac{1}{96}$ & $\tfrac{-1}{96}$ & $\tfrac{27}{5120}$ & $\tfrac{-13}{7680}$ & $\tfrac{ 11}{28672 }$ & $\tfrac{ -113}{ 1720320}$ & $\tfrac{ 1891}{ 212336640}$ & $\tfrac{ -3679}{ 3715891200}$ & $\tfrac{421}{4 541 644 800}$\\
  $n$=5  & 0 & $\tfrac{2}{375}$ & $\tfrac{-2}{375}$ & $\tfrac{42}{15625}$ & $\tfrac{-8}{9375}$ & $\tfrac{ 1556}{ 8203125}$ & $\tfrac{ -1294}{ 41015625}$ & $\tfrac{ 3254}{ 791 015 625}$ & $\tfrac{ -60 272}{ 138 427 734 375}$ & $\tfrac{162 556}{4 229 736 328 125}$\\
  $n$=6  & 0 & $\tfrac{1}{324}$ & $\tfrac{-1}{324}$ & $\tfrac{181}{116 640}$ & $\tfrac{-257}{524 880}$ & $\tfrac{ 89}{ 826 686}$ & $\tfrac{ -1553}{ 88 179 840}$ & $\tfrac{ 9161}{4 081 466 880 }$ & $\tfrac{ -74 143 }{ 321 415516 800}$ & $\tfrac{92 579}{4 714 094 246 400}$\\
  $n$=7  & 0 & $\tfrac{2}{1029}$ & $\tfrac{-2}{1029}$ & $\tfrac{82}{84 035}$ & $\tfrac{-232}{756 315}$ & $\tfrac{ 3476}{ 51 883 209}$ & $\tfrac{-134 }{ 12 353 145 }$ & $\tfrac{ 155 894}{ 114 402 475 845}$ & $\tfrac{ -550 288}{ 4 004 086 654 575}$ & $\tfrac{391 516}{34257185822475}$\\
  \end{tabular}
\end{table*}
Evaluating the terms of Eq.\ (\ref{eq:energyseries}) using the test functions described above and taking $\epsilon \rightarrow 0$ at the end yields the anticipated expansion for $\delta E_{2S}$ in moments,
\begin{equation}
  \begin{split}
    \delta E_{2S}/ \alpha  = & \frac{ \langle r_p^2 \rangle }{12 a_{\ell}^3 } - \frac{ \langle r_p^3 \rangle }{12 a_{\ell}^4 } + \frac{ 7 \langle r_p^4 \rangle }{160 a_{\ell}^5 }\\
    & - \frac{ 11 \langle r_p^5 \rangle }{720 a_{\ell}^6 } + \frac{ \langle r_p^6 \rangle }{252 a_{\ell}^7 } - \cdots .
  \end{split}
  \label{eq:momentseries2ndderivation}
\end{equation}
Thus, taking the ratio of the first two terms, we have the general result 
\begin{equation}
  \xi_{nS}(\rho , \ell ) = \frac{- \langle r_p^3 \rangle }{ a_\ell \langle r_p^2 \rangle } \sim \frac{- r_p }{ a_\ell },
  \label{eq:generalNLOsize}
\end{equation}
as long as $\rho$ has a range much smaller than the Bohr radius $a_\ell $.
Note that the generalization from $2S$ to $nS$ is inferred from Table \ref{tab:momentcoeffs}.
We are now able to claim---without any assumptions about the specific shape of the proton charge density or, equivalently, the form factor---that the inclusion of higher-order moments in Eq.\ (\ref{eq:LOcorrection}) is of negligible impact on the value of $\langle r_p^2 \rangle $ inferred from measurements of the $2S$-$2P$ Lamb shift.
\section{\label{sec:momentdep} Moment dependence of the finite-size correction for $n\neq 2$}
Writing the finite-size correction Eq.\ (\ref{eq:energyseries}) in the form
\begin{equation}
  \delta E_{nS} = \frac{\alpha }{a_\ell } \sum_{j=1}^\infty c_n^{(j)} \frac{ \langle r_p^j \rangle }{a_{\ell}^j } ,
  \label{eq:coefficientdefn}
\end{equation}
we can summarize the dependence of $E_{nS}$ on moments of the proton charge distribution via the dimensionless coefficients $c_n^{(j)}$.
In Table \ref{tab:momentcoeffs}, we list the coefficients for $n=1,\ldots ,7$ and $j=1,\dots ,10$ as calculated by the method introduced in the previous section.
The tabulated coefficients suggest that $c_n^{(3)} = -c_n^{(2)}$ for all $n$, so that Eq.\ (\ref{eq:generalNLOsize}) holds for any $S$-state.
\vspace{0cm}
\section{Conclusion}
We have shown that the proposed parametrization of Eq.\ (\ref{eq:gluckdeltaE}-\ref{eq:gluckq2}) does not explain the discrepancy in proton radius measurements obtained from electronic versus muonic hydrogen.
Using a typical approximation to the proton form factor as well as a more general coordinate-space analysis, we have determined the size of the effects of higher-order moments of the proton charge distribution.
The expansion of Eq.\ (\ref{eq:coefficientdefn}) combined with the coefficients in Table \ref{tab:momentcoeffs} could also be used for nuclear targets for which the sizes of the moments are more significant, as well as for transitions involving higher $S$-states in light atoms.
Our calculations Eq.\ (\ref{eq:sizeratio}) and Eq.\ (\ref{eq:generalNLOsize}) show that, while it is true that the next term in a series expansion for the finite-size correction in powers of $r_p / a_\ell$ is more important with muonic hydrogen, the correction cannot be the source of a 4\% discrepancy in $r_p $.
Furthermore, this term is not $\mathcal{O}(\langle r_p^4 \rangle )$, but is $\mathcal{O}( \langle r_p^3 \rangle )$.
The finite-size correction to $E_{nS}$ depends quite generally on all moments of the proton charge distribution of at least second-order.
This unexpected result has been verified using several different approaches.
We close with the reaffirmation that truncating the finite-size energy correction Eq.\ (\ref{eq:LOcorrection}) at $ \langle r_p^2 \rangle $ can make a difference of no greater than a few parts in $10^3$ for $\mu p$ and that the proton radius puzzle remains open in this regard.
\section{Acknowledgments}
This work was partially supported by the Seattle Chapter of the Achievement Rewards for College Scientists Foundation, and the U.S. Department of Energy Office of Science, Office of Nuclear Physics under Award Number DE-FG02-97ER-41014.
\vspace{0cm}
\bibliography {ref}

\begin{thebibliography}{10}

\bibitem{RevModPhys.84.1527}
Peter~J. Mohr, Barry~N. Taylor, and David~B. Newell.
\newblock Codata recommended values of the fundamental physical constants:
  2010*.
\newblock {\em Rev. Mod. Phys.}, 84:1527--1605, Nov 2012.

\bibitem{Pohl:2010}
R.~Pohl, A.~Antognini, F.~Nez, F.~D. Amaro, F.~Biraben, J.~M.~R. Cardoso, D.~S.
  Covita, A.~Dax, S.~Dhawan, L.~M.~P. Fernandes, A.~Giesen, T.~Graf, T.~W.
  Hansch, P.~Indelicato, L.~Julien, C.~Y. Kao, P.~Knowles, E.~O. Le~Bigot,
  Y.~W. Liu, J.~A.~M. Lopes, L.~Ludhova, C.~M.~B. Monteiro, F.~Mulhauser,
  T.~Nebel, P.~Rabinowitz, J.~M.~F. dos Santos, L.~A. Schaller, K.~Schuhmann,
  C.~Schwob, D.~Taqqu, Jfca Veloso, and F.~Kottmann.
\newblock The size of the proton.
\newblock {\em Nature}, 466(7303):213--216, 2010.

\bibitem{Antognini25012013}
Aldo Antognini, François Nez, Karsten Schuhmann, Fernando~D. Amaro, François
  Biraben, João M.~R. Cardoso, Daniel~S. Covita, Andreas Dax, Satish Dhawan,
  Marc Diepold, Luis M.~P. Fernandes, Adolf Giesen, Andrea~L. Gouvea, Thomas
  Graf, Theodor~W. Hänsch, Paul Indelicato, Lucile Julien, Cheng-Yang Kao,
  Paul Knowles, Franz Kottmann, Eric-Olivier Le~Bigot, Yi-Wei Liu, José A.~M.
  Lopes, Livia Ludhova, Cristina M.~B. Monteiro, Françoise Mulhauser, Tobias
  Nebel, Paul Rabinowitz, Joaquim M.~F. dos Santos, Lukas~A. Schaller,
  Catherine Schwob, David Taqqu, João F. C.~A. Veloso, Jan Vogelsang, and
  Randolf Pohl.
\newblock Proton structure from the measurement of 2s-2p transition frequencies
  of muonic hydrogen.
\newblock {\em Science}, 339(6118):417--420, 2013.

\bibitem{Pohl:2013yb}
Randolf Pohl, Ronald Gilman, Gerald~A. Miller, and Krzysztof Pachucki.
\newblock {Muonic hydrogen and the proton radius puzzle}.
\newblock {\em Ann. Rev. Nucl. Part. Sci.}, 63:175--204, 2013.

\bibitem{Carlson:2015jba}
Carl~E. Carlson.
\newblock {The Proton Radius Puzzle}.
\newblock {\em Prog. Part. Nucl. Phys.}, 82:59--77, 2015.

\bibitem{Gluck:2014qxa}
M.~Gl{\"u}ck.
\newblock {On the Proton charge extensions}.
\newblock 2014.
\newblock arXiv:1412.3423.

\bibitem{Carroll:2011de}
J.~D. Carroll, A.~W. Thomas, J.~Rafelski, and G.~A. Miller.
\newblock {Proton form-factor dependence of the finite-size correction to the
  Lamb shift in muonic hydrogen}.
\newblock 2011.
\newblock arXiv:1108.2541.

\bibitem{Carroll:2011rv}
J.~D. Carroll, A.~W. Thomas, J.~Rafelski, and G.~A. Miller.
\newblock {Non-Perturbative Relativistic Calculation of the Muonic Hydrogen
  Spectrum}.
\newblock {\em Phys. Rev.}, A84:012506, 2011.

\bibitem{Friar:2005jz}
James~Lewis Friar and I.~Sick.
\newblock {Muonic hydrogen and the third Zemach moment}.
\newblock {\em Phys. Rev.}, A72:040502, 2005.

\bibitem{Cloet:2010qa}
Ian~C. Cloet and Gerald~A. Miller.
\newblock {Third Zemach Moment of the Proton}.
\newblock {\em Phys. Rev.}, C83:012201, 2011.

\end{thebibliography}
\bibliographystyle{unsrt}
\end{document}